\documentclass{article} 
\usepackage{nips15submit_e,times}
\usepackage{hyperref}
\usepackage{url}
\usepackage{graphicx}

\title{Intelligence-based Cybersecurity Awareness Training - an Exploratory Project}
\author{
Tam n. Nguyen, Lydia Sbityakov, Samantha Scoggins\\
North Carolina State University\\
\texttt{\{tam.nguyen, lesbitya, smscoggi\}@ncsu.edu} \\
}

\nipsfinalcopy 
\begin{document}
\maketitle

\begin{abstract}
  Cybersecurity training should be adaptable to evolving the cyber threat landscape, cost effective and integrated well with other enterprise management components. Unfortunately, very few cybersecurity training platforms can satisfy such requirements. This paper proposes a new and novel model for conducting cybersecurity training with three main objectives: (i) training should be initiated by emerging relevant threats and delivered first to the most vulnerable members (ii) the process has to be agile (iii) training results must be able to provide actionable intelligence. For the first time, this paper establishes a type system (ontology and associated relationships) that links the domain of cybersecurity awareness training with that of cyber threat intelligence. Powered by IBM Watson Knowledge Studio platform, the proposed method was found to be practical and scalable. Main contributions such as exports of the type system, the manually annotated corpus of 100 threat reports and 127 cybersecurity assessment results, the dictionaries for pre-annotation, etc were made publicly available.
\end{abstract}

\section{Introduction}
The threat landscape is constantly changing. Something that was not considered a vulnerability yesterday may now become one \cite{Manadhata2011AnMetric}. Therefore, cybersecurity awareness training (CAT) should be adaptable, cost effective and, most importantly, compatible to integrate well with other components. Examples of other components for integration would be enterprise risk management, incident management, threat intelligence and so on. Unfortunately, in most cases, CAT is not a strong component in most cyber defense strategy \cite{Jakoubi2009AManagement}.

This lack of emphasis on human training and analysis leads to greater issues with establishing cybersecurity requirements, cyber incident's impact determination, and the simulation of possible attack scenarios. For example, security requirements tend to be mechanical 1-on-1 mappings from obvious security features and regulatory controls \cite{Cleland-Huang2014HowGratae}. Consequently, there is a very common assumption that most malicious hackers will seek the path of the most devastating exploits. In reality, hackers do not follow a straight line while using only a fraction of available vulnerabilities to deliver attacks \cite{Allodi2017TowardsAssumptions}. A lot of those vulnerabilities deal directly with human errors \cite{Messaoud2017AdvancedChallenges}.

Also, it is found that when evaluating cybersecurity risks, there are fixation on binary events without consideration of fuzzy states in between, and bias toward the point of view of security management (more technical) rather than overall business goals (more human oriented) \cite{Dhillon2011Developer-drivenTrenches} \cite{Bayuk2013SecurityConstruct}. Last but not least, common reported threat metrics can be highly debatable among teams and agencies \footnote{https://bit.ly/1yJcGjC} since numbers cannot describe all possible underlining context. 

This paper proposes InCAT - a new model for conducting cybersecurity training with a strong focus on drilling deep into the shared contexts among collected cyber awareness training results, cyber threat intelligence reports, and other cybersecurity related data logs. InCAT stands for "Intelligence-based Cybersecurity Awareness Training" and its feedback loop starts with a threat intelligence feed where the most recent cyber threats will be analyzed by machine learning models to identify the current attack-defense themes. This angle is called "Technical Angle". From the identified themes, quizzes will be sent to users (samples) within a company (population). Machine learning models will analyze users' responses with expected results such as a list of vulnerabilities for which employees are least prepared for. This angle is called "Human Angle". Actionable intelligence can then be derived from analyzing results gathered from both angles.

The main contributions of this paper include: (i) A novel new model for conducting cybersecurity awareness training that is highly adaptive to threat landscape (ii) Downloadable type system, dictionaries, and human annotated data-sets for further customization and studies (iii) Exported machine learning models and starter code base for immediate deployment.

\section{Backgrounds}
Humans are the weakest link in any cyber defense strategy. 78\% of cyber incidents were caused by careless humans\cite{PonemonInstitute20172017Overview}. The process of cyber awareness training is full of challenges. First, the threat landscape is evolving rapidly with both internal factors (technology changes, business flows changed, etc) and external factors (changes in supply chain, compliance, competitors, enemies, political climates, etc) \cite{Ingalsbe2008ThreatEnd, Manadhata2011AnMetric}. To make things worse, there is no agile cooperation between cyber awareness education and other departments. 

Second, knowledge is not always translated into correct actions. For example, people who know the types of phishing are not completely immune from actual phishing. Training materials are more focused on teaching knowledge rather than the skill. Consequently, tests are developed to test just the knowledge. In addition, learners are well aware that they are being tested and are guarded in their answers. It is an arduous task to simulate real world scenarios. Finally, it is challenging to prepare people for potential unknown threats from creative adversaries that may be state sponsored and have not happened yet.

Abawajy \cite{Abawajy2014UserMethods} did research on user preferences of cybersecurity awareness methods. Even with a "low-tech" appearance, the performance of text-based education is on par with other "fancy" methods. Pawlowski \cite{Pawlowski2016SocialDesign} presented a way to map learners' perceptions toward cybersecurity topics using data analytics. It was discovered that learners who have been exposed to several cybersecurity key terms and concepts, care very much about cybersecurity topics that may affect them personally. However, they lack awareness of the bigger picture which also includes national cyber infrastructure and cyber terrorism.

Wei \cite{Wei2017IntegratingAssessment} adopts concept mapping (CM) as a tool to enhance the teaching and evaluation of Information System courses. By analyzing the topology of student CMs, instructors can design CM-based tasks, and grade student CMs against a master CM. Green \cite{Green2018TowardsSchemes} proposed an alternative to regular text-based machine learning approach with heavier emphasis on argument mining. The process involves steps of careful annotation of a selected corpus, building logic rules or relationships based on domain specific knowledge, and finally, inferring arguments. The paper made a case for a more sophisticated content analysis method when dealing with complex contents such as biological research papers.

Joshi \cite{Joshi2013ExtractingText} described a strategy of "combining" intelligence gathered from both structured and unstructured texts to form actionable intelligence in the domain of cybersecurity. Ferruci \cite{Ferrucci2010BuildingProject} wrote about IBM's billion dollar baby - Watson, and proves that within a large knowledge domain such as Jeopardy's, not only deep learning can understand human languages and response with thoughtful insights, but it can also perform reliably in real time. To sum up the strength and weaknesses of Watson, Deloitte provided a good report \cite{DeloitteDevelopment2015DisruptionWatson} in 2015. At the time of this paper, IBM Watson has been deployed commercially in various domains including health care, finance, insurance and so on.

\section{Methodology and Designs}
\begin{figure}[h]
  \centering
  \includegraphics[width=11cm]{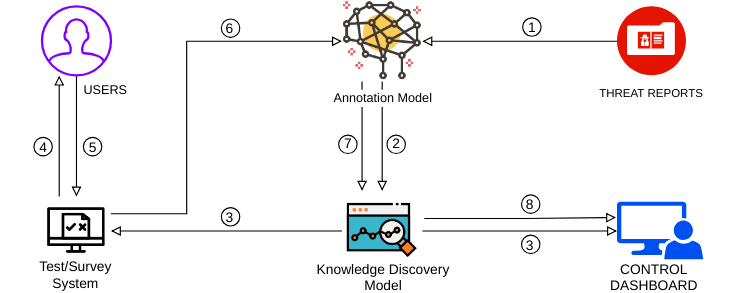}
  \caption{INCAT Design Structure}
  \label{Figure:IncatDesign}
\end{figure}

InCat feedback loop includes 8 main steps as shown in Figure \ref{Figure:IncatDesign}. 

Threat reports are highly condensed data that relate to cyber threat developments and was prepared by domain experts. Reports will be analyzed by the Annotation Model for identification of key entities and the relationships between them. Newly annotated data will be further analyzed by the Knowledge Discovery Model which will automatically identify newly emerged patterns in cyber threat landscape. The Test/Survey System will construct assessments based on the knowledge components relating to the newly identified patterns and deliver to the users who are most vulnerable. Assessment results will then be returned to the Test/Survey System for initial processing, resulting in reports with scores and users' response texts. Such user assessment reports will be analyzed by the same Annotation Model and the Knowledge Discovery Model. Derived knowledge from threat reports (step 3) and user assessment reports (step 7) are stored in database and be displayed by a Control Dashboard system. Through this dashboard, cybersecurity analysts may correlate details and derive further actionable intelligence.

\begin{figure}[h]
  \centering
  \includegraphics[width=12cm]{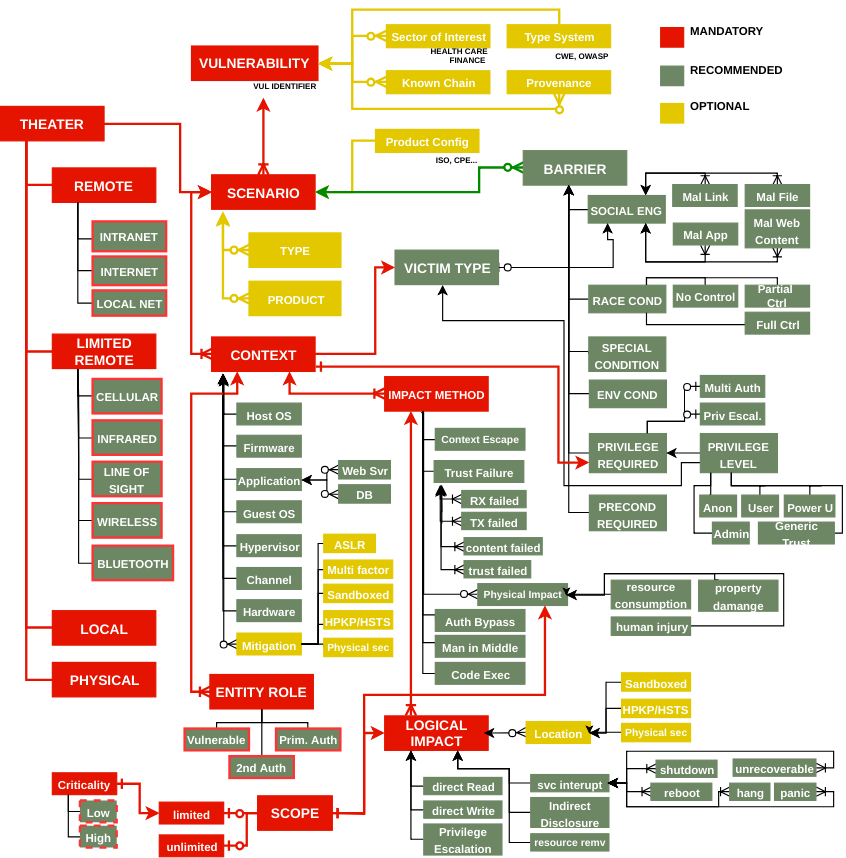}
  \caption{Vulnerability Description Ontology}
  \label{Figure:VulOntology}
\end{figure}

There are three main steps for building the Annotation Model: (i) Building a system type (ii) Building ground truths by careful manual annotation (iii) Training and evaluating models. A system type is a domain-specific ontology enriched with relationships. The ontology serves as the core of a common language describing cyber security vulnerabilities that all models within the InCAT system will use. A well-designed ontology is crucial in finding actionable intelligence across domains, in highly complicated situations such as court cases \cite{Michel2018CyberCybercrime}. However, there are at least two big problems with existing cyber security ontologies: incomplete and incompatible \cite{Mavroeidis2017CyberIntelligence}. The paper decided to rely on the NIST's recommendations for Cyber Vulnerability Description ontology \cite{Booth2016DraftOntology} which was summarized into Figure \ref{Figure:VulOntology}. While being not perfect, NIST's ontology for Cyber Vulnerability Description appears to be the latest attempt with continuing development efforts sponsored by the US Government.
\section{Implementation}
\subsection{Data Source and Preprocessing}


The National Vulnerability Database (NVD) \footnote{https://nvd.nist.gov/vuln/data-feeds} was already a high quality data set so minimal manipulation of the data was needed for preprocessing. Our source was the 2018 vulnerability .json file, CVE-2018, downloaded on September 19th (The file is continually updated). We then created two data sets one for the text analysis and the second for clustering analysis.  

For text analysis we extracted the CVE-ID number and Description for 8748 records.  For the categorical data we choose to work with the more updated Base Metrics 3.0 (BM3). The categorical data set used for clustering consists of all 6851 available 2018 NVD records with a completed BM3 section.  This data was not available for every record since it only becomes available after a vulnerability analysis is completed; records without completed BM3 fields were eliminated from clustering analysis.  BM3 features generally correspond to the vulnerability ontology used in the text analysis (see next section) which is based on the NIST's recommendations for Cyber Vulnerability Description ontology \cite{Booth2016DraftOntology}.

To cluster the data we extracted the CVE-ID and categorical fields (\ref{table:categoricalData}).  Derivative fields such as \textit{Base Score} and \textit{Base Severity} that are calculated from other fields and the fields product and vendor were eliminated from cluster processing.

\subsection{Identifying Attack vectors}
\begin{table}[h!]
\begin{center}
\begin{tabular}{ |l|l| } \hline
Feature & Values\\\hline
Attack Vector & Network, Adjacent, Local, Physical  \\ 
Attack Complexity & Low, High  \\ 
Privileges Required & None, Low, High  \\ 
User Interaction & None, Required  \\ 
Confidentiality Impact & High, Low, None\\
Integrity Impact & High, Low, None\\
Availability Impact & High, Low, None\\
\hline
\end{tabular}
\end{center}
\caption{Categorical Data} 
\label{table:categoricalData}
\end{table} 

This relatively small set of features and possible values can have 1,296 different combinations.  However, a simple data query reveals that only 236 combinations were associated with any threats and 74.7\% of the vulnerabilities fall into just 16 unique combinations of features.  Finding meaningful clusters in the data would help to focus training on the most prevalent threat vectors and make the most efficient use of resources. Because the data is purely categorical common algorithms such as k-means are not recommended. \cite{Tan2019IntroductionMining}.  DBScan while helpful for removing outliers is not recommended for data with higher dimensions where Euclidean distance is less effective.  The KModes module available for python was selected for the clustering implementation \footnote{https://pypi.org/project/kmodes/}  because this package is readily available and specifically designed to work with categorical data.

KMode defines clusters based on the number of matching categories between data points and it calculates the cost or dissimilarity of each set of clusters based on the sum of number of fields that are different between each record and its assigned centroid.  When validating the hyper-parameters dozens of combinations of centroid initialization algorithms and the number of clusters were tried.  The possible range of number of clusters for this dataset is 1-236, however a more reasonable number of clusters would be in the range of 5-20.  The combination of the Huang initialization algorithm and ten clusters were selected for our final clustering based on the elbow in the graph of the validation data.  Because there is an element of randomness in the Huang algorithm the same exact results are not necessarily repeated in subsequent iterations.

The cost of our final clusters ended up being only 4473 which means that on average records differed from their assigned centroid by less than 1.  This shows that the data is naturally strongly clustered and makes it likely that targeting specific features for training would be effective.  Specifically, note that nearly all of the clusters indicate NETWORK for ATTACK VECTOR meaning that being on a local network or physical access to the machine is not typically a risk factor for these types of vulnerabilities.  In addition, almost half of the clusters indicate REQUIRED for USER ACTION.  Certainly in these cases risk would be mitigated by educating users about the specific actions NOT to take when faced with one of the associated vulnerabilities. \footnote{Cluster code and data can be found at https://github.com/genterist/INCAT-Project}

\begin{figure}[ht]
  \centering
  \includegraphics[width=8cm]{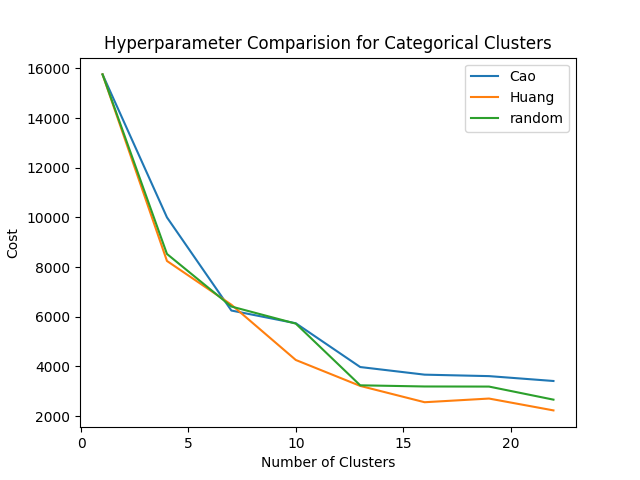}
  \caption{Hyperparameter Comparison}
\end{figure}

\begin{table}[ht]
\begin{center}
\begin{tabular}{ |l|p{1.3cm}|p{1.8cm}|p{1cm}|p{1.3cm}|p{1cm}|p{1cm}|p{0.7cm}| } \hline
Attack Vector & Comple- xity & User Interaction & Privi- leges & Confiden- tiality & Integrity & Availa- bility & Count \\ \hline
NETWORK & LOW & NONE & NONE & HIGH & HIGH & HIGH & 2394 \\
NETWORK & LOW & NONE & NONE & NONE & HIGH & NONE & 853 \\
NETWORK & LOW & NONE & NONE & HIGH & NONE & NONE & 711 \\
NETWORK & LOW & NONE & NONE & NONE & NONE & HIGH & 656 \\
NETWORK & LOW & REQUIRED & NONE & LOW & LOW & NONE & 638 \\ 
LOCAL & LOW & NONE & LOW & HIGH & HIGH & HIGH & 525 \\
NETWORK & LOW & REQUIRED & LOW & LOW & LOW & NONE & 486 \\
NETWORK & LOW & REQUIRED & NONE & NONE & NONE & HIGH & 296 \\
NETWORK & LOW & REQUIRED & NONE & HIGH & NONE & NONE & 211 \\
NETWORK & LOW & NONE & LOW & LOW & LOW & LOW & 81 \\
\hline
\end{tabular}
\caption{Optimal Clusters}
\end{center}
\label{table:1}
\end{table}

The clustering results serve to inform the annotation and classification process used in 4.3 and in future could provide the underlying framework for the identification and selection of Attack Vectors on an on-going basis.

\subsection{Annotation Model for Classifying Text Data}
Our datasets of descriptions for Threat Reports were extracted from the National Vulnerability Database which can be considered the gold standard of threat reporting and is being used nation wide. In the form of XML or CSV files, which are then transformed into JSON files, datasets can be imported into Annotation Model which is powered by the IBM Watson Knowledge Studio platform \footnote{https://www.ibm.com/watson/services/knowledge-studio/}.

It is notable that in a real-world scenario, reports will only touch a few "boxes" of this crowded ontology. Therefore, the ontology has to be enriched with relationships that are context-preserving. Such relationships can be described directly by rules or indirectly by type \& sub-type specifications. The novelty of our work comes in the form of our own designed relationship rules, the overloading of the rules and the re-mapped hierarchy of the original NIST's ontology. The detailed description file of our type system (ontology with relationships) can be viewed at our Github repository \footnote{https://github.com/genterist/INCAT-Project}.

Human annotators must manually annotate documents to establish ground truths.  For our first approach small sets of 50 entries each were extracted from our corpus of ~8000 NVD threat reports in 2018.  Two team members worked on their corresponding assigned non-overlapping sets and manually annotate entities together with relationships. The team member who did not do any manual annotation reviewed the annotations and published annotated entries as ground truths.  

On the IBM Watson Knowledge Studio platform, we trained and tested our model using the manually annotated entries which are separated into training set (70\%), test set(23\%), and blind set(7\%). Training set is used to teach machine the domain specific knowledge through annotated entities and their relationships. Trained model will then perform on test set to produce test set machine results. Upon comparing test set machine results with human annotated results, the accuracy of the model can be defined. Blind set is used to test the system only after several iterations of training and testing. The end result of this process is an annotation model that can be deployed with other models which, in our case, happened to be IBM Knowledge Discovery.

Our original annotation approach for vulnerability descriptions resulted in a 11\% success rate for the model so we started over using a more standard approach using the Watson platform's ability to randomly assign overlapping documents to a pair of individual annotators.  Small sets of 50 randomly selected documents were assigned to two annotators with a 50\% overlap.  Conflicts were resolved on the platform and promoted to ground truth for training.  Agreement between the annotators was calculated for entities, relationships, and co-references.  In the first round our agreement on entity mentions ranged from 0 to 1 with an overall score of .61, already a huge improvement.

\section{Analysis}
Some of the challenges encountered by the human annotators on this project included the inflexibility and ambiguity in certain areas of the original ontology.  In addition, certain entities showed up very rarely in the corpus making training for them difficult.  Frequently, information was gleaned from the descriptions that there was no way to annotate either unambiguously or at all.  A more flexible ontology that could incorporate features of natural grammar would make this a much more powerful tool.  Making the overall model less hierarchical and adding additional roles and context clues would be a great improvement.

Natural language inherently has ambiguities that cause serious conflicts with the precise nature of describing cyber security scenarios. For example, "Windows" may mean "Operating system", or "Desktop". Therefore, even when the training sets were well annotated, the model will have issues with learning certain relationships due to the lack of samples. The model's accuracy will go down when more annotated content about new scenarios were added, diluting or even confusing prior knowledge about known scenarios. This means training should better be done in a rolling basis.

While IBM Watson is a commercial product with verified natural language processing capabilities,it is important to note that machine learning models learn from what was shown and cannot learn from the details that they have not yet seen. The accuracy also depends on the type system being used. While NIST has excellent credibility, their ontology is still being developed. There are also some ontology tweaking required for deployment with IBM Watson platform. Our type system is therefore experimental.

There are generally two approaches for this kind of tasks: using machine learning (such as IBM Watson) or using logic programming (Prolog). Relationship documentation is the heaviest task in the data preparation phase. IBM Watson provides a convenient GUI, point-and-click interface for annotating relationships, while Prolog requires coding of relationships. With Prolog, a domain expert has to know both the domain knowledge and Prolog programming skills. In that sense, IBM Watson is more scalable than Prolog.

However, the corpus has to be large enough (thousands of entries) in order for Watson to learn relationships accurately. With Prolog, logics that are known can be programmed by the domain experts without relying too much on training data, which appears to be faster and more accurate than Watson. However, Prolog cannot infer logics that are not built on the existing logics stored in its knowledge base (closed world). Machine learning models like Watson, on the other hand, was designed to spot new patterns that even human experts are not aware of as long as there are enough annotated data-sets. We believe there is also a possibility of integrating rule-based logic into our model for faster and more efficient performance.

\section{Conclusion}
InCAT is a new model for conducting cybersecurity training with three long term goals: (i) training efforts are initiated by emerging relevant threats and delivered first to the most vulnerable groups (ii) each training session must be promptly executed (iii) training results must be able to provide actionable intelligence to be employed by other systems such as enterprise risk management, enterprise threat intelligence, etc.

The project is the first in making an attempt at constructing a type system that allows the inter-connecting of intelligence among cybersecurity related domains, including cybersecurity awareness training. The project is also the first in designing an intelligence-driven rather than compliance-driven or knowledge-driven process for conducting cybersecurity awareness training. Within a very short amount of time, the project was able to provide preliminary results including a manually annotated corpus of 100 threat reports, and 127 cybersecurity awareness training assessments; a set of dictionaries for pre-annotation; and an annotation model capable of 66\% entity marking coverage with 75\% accuracy.

Basic research questions were answered and directions for future developments are clear: (1) Deeper research into inter-domain ontology design and upgrade the type system, (2) Significantly extend the words in dictionaries for more comprehensive pre-annotation (3) Integrate Prolog into work flow, allowing it to work in parallel with the machine learning model (4) Extend the corpus with more human annotated contents.

\bibliographystyle{IEEEtran}
\bibliography{IEEEabrv,references.bib}

\end{document}